\documentclass[conference]{IEEEtran}
\usepackage{xcolor}
\usepackage[T1]{fontenc}
\usepackage{cite}
\usepackage{amsmath,amssymb,amsfonts}
\usepackage{graphicx}
\usepackage{textcomp}
\usepackage{xcolor}
\usepackage{enumitem}
\usepackage{xspace}
\usepackage{tabularx,colortbl}
\usepackage{multirow}
\usepackage{booktabs}
\usepackage{hyperref}

\newcommand{\xray}{x-ray\xspace}
\newcommand{\verbal}{verbal\xspace}
\newcommand{\arrow}{arrow\xspace}
\newcommand{\noai}{no-AI\xspace}
\providecommand{\Description}[1]{}

\hypersetup{pdfauthor={Omar Elgohary, Amanda Elvarsdottir, Vannessa Lee, Zhu-Tian Chen}}

\begin{document}

\title{How Contextual Information in Physical Environments Shapes Decisions with AI}

\author{
\IEEEauthorblockN{1\textsuperscript{st} Omar Elgohary}
\IEEEauthorblockA{\textit{University of Minnesota}\\
\textit{Twin Cities}\\
elgoh003@umn.edu}
\and
\IEEEauthorblockN{2\textsuperscript{nd} Amanda Elvarsdottir}
\IEEEauthorblockA{\textit{University of Minnesota}\\
\textit{Twin Cities}\\
elvar002@umn.edu}
\and
\IEEEauthorblockN{3\textsuperscript{rd} Vannessa Lee}
\IEEEauthorblockA{\textit{University of Minnesota}\\
\textit{Twin Cities}\\
jiang166@umn.edu}
\and
\IEEEauthorblockN{4\textsuperscript{th} Zhu-Tian Chen}
\IEEEauthorblockA{\textit{University of Minnesota}\\
\textit{Twin Cities}\\
ztchen@umn.edu}
}

\maketitle

\begin{abstract}
AI assistants in augmented reality may advise users about parts of the physical world they cannot directly observe. We ask whether contextual information from the situation helps users make better decisions with fallible AI. In an AR visual-search study ($N=32$), we varied target prevalence and scene visibility while participants judged whether a key was present, with or without AI assistance. Lowering target prevalence and revealing part of the desk together improved observed AI-assisted accuracy, but our analysis did not demonstrate that this gain came from participants making better use of context. Context also changed which answers seemed plausible: when keys were rare, participants became more skeptical of \emph{Present} suggestions, but this shift applied to both correct and incorrect suggestions rather than improving selective acceptance of correct advice. Revealing part of the desk also increased decision time, with little evidence of better AI use. Overall, contextual information can shape AI-assisted decisions and increase decision effort without necessarily helping users judge when the AI is correct.
\end{abstract}


\begin{figure*}[t]
  \centering
  \includegraphics[width=0.92\textwidth]{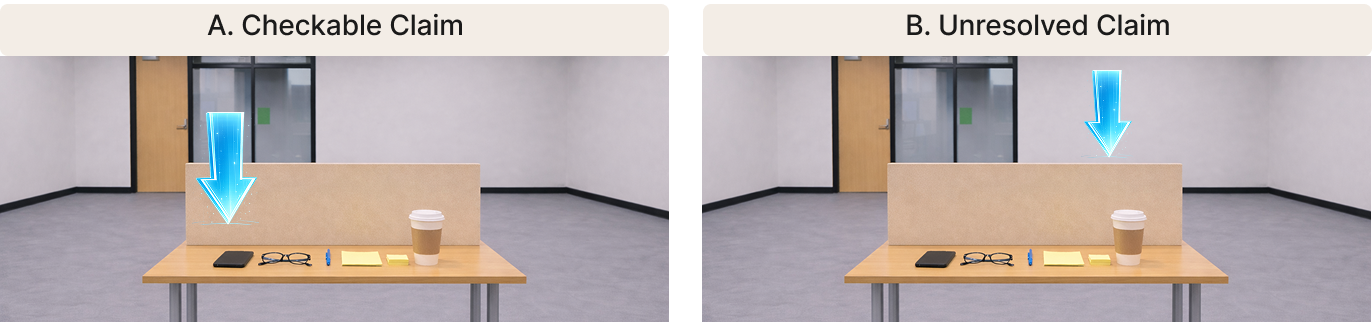}
  \caption{The same in-scene AI cue can support different judgments depending on evidence access. (A) When the indicated region is visible, participants can directly check the spatial claim; an empty region does not by itself establish desk-wide absence. (B) When the indicated location lies behind the occluder, contextual information may change plausibility but cannot verify the claim.}
  \Description{Two AR screenshots compare an arrow pointing into a visible region with an arrow pointing behind an occluding divider. The visible-region claim can be checked directly; the hidden-region claim remains unresolved.}
  \label{fig:teaser}
\end{figure*}

\section{Introduction}
\label{sec:intro}

AI assistants increasingly support decisions about physical environments that users cannot fully observe. AI glasses such as Ray-Ban Meta bring assistance into everyday settings~\cite{Meta2024RayBanAI}, while AR systems support room-accessibility inspection~\cite{Su2024RASSAR}, situated explanations~\cite{Xu2023XAIR}, and everyday actions~\cite{Cho2025PersistentAssistant,Li2024OmniActions}. These developments make a longstanding human--AI question consequential in situated interaction: when does AI advice help people make better decisions~\cite{Parasuraman1997Automation,Lee2004TrustAutomation,Schemmer2023AppropriateReliance,Fok2024Verifiability}? Users may accept incorrect advice or reject correct advice, motivating work on appropriate reliance~\cite{Zhang2020TrustCalibration,Bansal2021WholeExceedParts,Bucinca2021TrustThink,He2023KnowingAboutKnowing,Klingbeil2024TrustReliance}.

Consider an assistant claiming that a key lies behind an occluder (Fig.~\ref{fig:teaser}B). A user may know that keys are uncommon~\cite{Wolfe2007LowPrevalence}, or inspect part of the desk and find no key~\cite{Manavalan2026VisibilityPrevalence}. Both inform the probability of a hidden key without establishing whether it exists. Following Dey's broad definition of context as situational information~\cite{Dey2001Context}, we use \emph{contextual information} more narrowly for information available independently of the AI that changes an outcome's plausibility without directly resolving it.

Prior work examines explanations, uncertainty displays, and cognitive forcing~\cite{Zhang2020TrustCalibration,Bansal2021WholeExceedParts,Wang2021ExplanationsHelpful,Fok2024Verifiability,Papenmeier2022ComplicatedTrust,Cao2024AppropriateReliance,Bucinca2021TrustThink}. We examine independently available physical context when the relevant state and the AI-referenced location remain hidden. Such context may change which answers users accept without identifying individual AI errors. For example, greater skepticism toward \emph{Present} suggestions can reject both correct and incorrect advice. We therefore distinguish decision accuracy, changes in response policy, and selective agreement with correct versus incorrect advice.

We conducted an AR visual-search study ($N=32$) crossing target prevalence (25\% or 50\%) and scene visibility (fully or half occluded), with and without simulated fallible AI advice. Our focal analyses retain trials where visible information cannot directly resolve the answer. The combined context increased observed AI-assisted accuracy, but an exploratory decomposition accounted for most of this increase through the target mix and retained AI suggestions. The \emph{stronger prior} (lower prevalence) shifted agreement with \emph{Present} suggestions, while the \emph{target-free half} increased decision time; we did not detect increased agreement selectivity on these unresolved trials.

Our contributions are a controlled study of these contextual signals alongside fallible AI in AR, an analysis separating observed accuracy changes from changes in participant responses, and implications for designing and evaluating situated assistance as access to evidence changes.

\section{Related Work}
\label{sec:related-work}

\subsection{Using Fallible AI Advice}

Research on appropriate reliance examines when people accept or reject AI advice~\cite{Schemmer2023AppropriateReliance,Fok2024Verifiability}. Explanations can change reliance without consistently improving error detection~\cite{Bansal2021WholeExceedParts,Wang2021ExplanationsHelpful}; cognitive forcing~\cite{Bucinca2021TrustThink}, uncertainty displays~\cite{Cao2024AppropriateReliance}, and beliefs about AI reliability and human competence also shape responses~\cite{Yin2019AccuracyTrust,He2023KnowingAboutKnowing,Ma2023WhoTrust}.

Fok and Weld distinguish \emph{verifiability} from understanding an AI's reasoning, and \emph{outcome-graded} reliance from a defensible \emph{strategy-graded} policy given available information~\cite{Fok2024Verifiability}. Thus, following an incorrect prediction need not imply an irrational strategy when correct and incorrect predictions are observationally indistinguishable. Inspectable, claim-relevant evidence can support error detection~\cite{Gonzalez2021DetectErrors,Vasconcelos2023ReduceOverreliance,Morrison2023ExplanationStrategies}. Building on this distinction, our study measures accuracy and agreement selectivity separately for independently available physical context.

\subsection{Contextual Information in Physical Environments}

Hemmer et al. show that information asymmetry can enable humans to use contextual information to adjust AI decisions and achieve complementary performance~\cite{Hemmer2022InformationAsymmetry}. Our question concerns a restricted case: a \emph{stronger prior} toward absence and a \emph{target-free half} inform the hidden state but do not directly reveal it. We examine whether these cues change response policy, agreement selectivity, and accuracy after accounting for the cases retained in analysis. This characterizes the unresolved-evidence setting within the broader problem of human--AI complementarity.

Visual-search research establishes that target prevalence changes present/absent response tendencies~\cite{Wolfe2007LowPrevalence,Wolfe2010VaryingPrevalence}. Manavalan et al. crossed prevalence with partial visibility and evaluate signal-detection, Bayesian, and probability-matching accounts in a non-AI task~\cite{Manavalan2026VisibilityPrevalence}. Their work motivates our manipulation; our contribution is its study alongside fallible AI advice in AR and the separation of response changes from changes in the scored trial distribution.

Reliance also occurs within broader reasoning practices~\cite{Kawakami2022ChildWelfare,Sivaraman2026ReasoningCues}. Spitzer et al. report benefits from contextual information about task and AI capabilities in delegation~\cite{Spitzer2025Delegation}; our contextual cues instead concern an unresolved physical state. Situated visualization embeds data in physical environments~\cite{Willett2017EmbeddedData,Lee2023DesignPatterns}. Closely related, Liu et al. compare AR x-ray with a 2D minimap and show how perceptual difficulties and realistic embedded representations can contribute to inappropriate AI reliance~\cite{Liu2026ARReliance}. We instead vary prevalence and scene visibility, pooling AI formats to examine contextual effects. Together, these studies motivate evaluating what environmental information allows users to infer or verify, beyond how an AI cue is displayed.

\section{Study Rationale}
\label{sec:rationale}

\subsection{Context and Decision Benchmarks}
\label{sec:rationale-context}

Participants judged whether a Unity-rendered key was present on a physical desk viewed through an AR headset. A virtual divider hid all or half of the desk, while blocks varied target prevalence (25\% or 50\%). We call 25\% prevalence the \emph{stronger prior} toward absence (lower prevalence), the visible half containing no key the \emph{target-free half}, and their combination the \emph{combined context}. Both cues inform target presence without revealing a hidden key~\cite{Manavalan2026VisibilityPrevalence}.

A useful decision policy maximizes expected accuracy given the information available, rather than rejecting every AI error in hindsight~\cite{Fok2024Verifiability}. For a one-trial benchmark using prevalence and the AI's binary suggestion, assume equal error costs and the programmed 80\% accuracy for each target state. With the desk fully hidden, always answering \emph{Absent} yields 50\% or 75\% expected accuracy at the two prevalence levels; always following the AI yields 80\% at either level. For a \emph{Present} suggestion and prevalence $p$, Bayes' rule gives
\begin{equation}
P(\mathrm{Present}\mid\mathrm{AI{+}})=
\frac{0.8p}{0.8p+0.2(1-p)}.
\end{equation}
This is 0.80 at 50\% prevalence and approximately 0.57 at 25\%; an \emph{Absent} suggestion favors absence in both conditions. Following the AI is therefore optimal for this limited-information, single-trial benchmark, even though prevalence changes posterior belief. The benchmark uses the programmed reliability, which participants were not told, and does not model learning or trial history.

This benchmark applies to fully hidden trials. Partial visibility changes the information available, and excluding directly checkable cases changes the evaluated trials. Policy comparisons must therefore use the same trials and account for the joint information in the scene and AI cue; 80\% is not a fixed reference for every analyzed subset.

\subsection{Outcomes and Research Questions}
\label{sec:rationale-theory}
\label{sec:rationale-rqs}

We distinguish three outcomes. \emph{Accuracy} measures final correctness. \emph{Agreement selectivity} is the difference in following correct and incorrect suggestions, assessed separately for \emph{Present} and \emph{Absent} advice. For RQ2, this distinguishes a general response tendency from agreement associated with advice correctness. Selectivity does not establish whether a policy is optimal given available evidence: when individual AI errors cannot be observed, a sound policy may follow both correct and incorrect suggestions. \emph{Decision time} measures the interval from trial onset to final response, with and without AI; it does not isolate inspection or effort. These outcomes motivate:

\begin{itemize}[noitemsep,leftmargin=*]
    \item \textbf{RQ1 (Accuracy).} Does contextual information increase observed AI-assisted accuracy, and is any gain associated with changed participant responses?
    \item \textbf{RQ2 (Responses to AI).} Does context change agreement selectivity or mainly shift which answers participants accept?
    \item \textbf{RQ3 (Decision time).} Does context change decision time, and does the change differ with and without AI?
\end{itemize}

\section{Study Design}
\label{sec:design}

We conducted a within-subjects AR study (N=32) in which participants judged
whether a Unity-rendered key was present on a physical desk viewed
through an AR headset, with or without fallible AI advice.
The full experiment crossed two contextual factors---target prevalence
and scene visibility---with four AI conditions.
Because the present paper focuses on contextual information, prevalence
and visibility are the focal manipulations; the primary AI-assisted
analyses pool across the three AI presentation formats.
The protocol was approved by our institutional review board, all
participants provided informed consent, and the study was piloted with
four participants.

\subsection{Task and Physical Environment}

\begin{figure}[t]
    \centering
    \includegraphics[width=0.8\columnwidth]{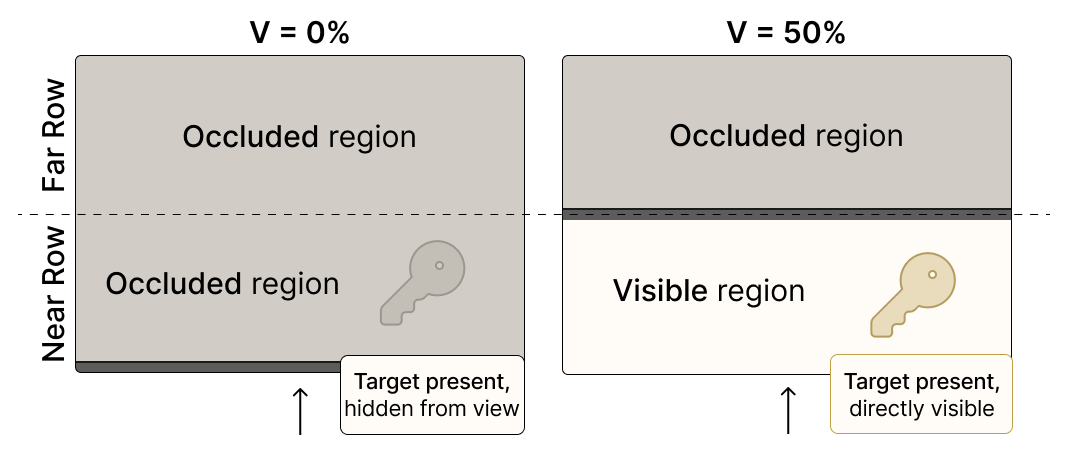}
    \caption{Visibility manipulation and analysis boundary for the Unity-rendered task target on the physical desk. At 0\% visibility (left), the desk was fully occluded. The 50\% example (right) shows a directly visible target, which was excluded from the focal analyses. Focal \emph{target-free half} trials used the same geometry but contained no key in the visible region, leaving the hidden half unresolved.}
    \label{fig:context-arrangement}
\end{figure}

\begin{figure*}
    \centering
    \includegraphics[width=\textwidth]{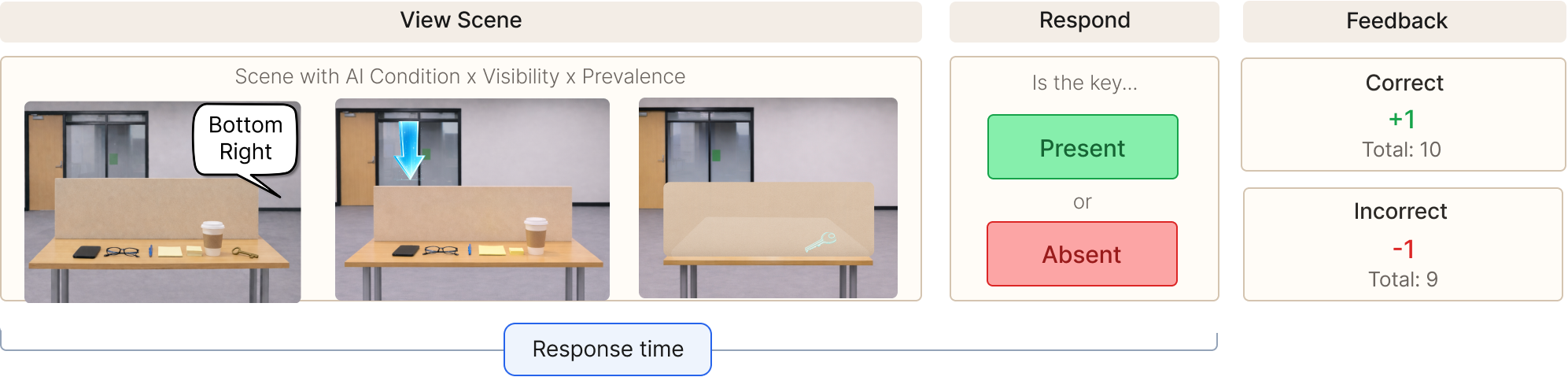}
    \caption{Examples of the three AI presentation formats and trial flow. Participants viewed the assigned context and AI cue, made a \emph{Present}/\emph{Absent} judgment using the response buttons, and received correctness and score feedback.}
    \label{fig:trial_process}
\end{figure*}

The desk was physical, while the task target was a Unity-rendered key
spatially registered to its surface. Target presence and position were
set by the trial state, separately from the AI suggestion. Participants
made a binary \emph{Present}/\emph{Absent} judgment about this task
target. When present, it appeared at a candidate location in the near
or far row; the virtual divider controlled its visibility. The task
target was distinct from the translucent key used to depict x-ray
advice.
Present targets were distributed evenly between the two rows, with
randomized horizontal position and spatial jitter.

Distractor objects in the visible region required participants to search
for the key before concluding that this region was target-free.

\subsection{Context Manipulations}

We manipulated two sources of contextual information: how likely the
key was to appear and how much of the desk participants could inspect.

\paragraph{Target prevalence}
We manipulated participants' prior expectation by changing how often
the key appeared.
In 25\% prevalence blocks, the key was present on 10 of 40 trials; in
50\% prevalence blocks, it was present on 20 of 40 trials.
The current prevalence was displayed throughout each block, and the
order of prevalence conditions was counterbalanced.

\paragraph{Scene visibility}
We manipulated partial observation using a virtual divider
(Fig.~\ref{fig:context-arrangement}).
At 0\% visibility, the entire desk was hidden.
At 50\% visibility, the near half of the desk was visible while the far
half remained hidden.
If the key appeared in the visible half, participants could directly
determine that it was present.
If the visible half contained no key, participants could rule out that
part of the desk but could not determine whether a key was hidden behind
the divider. Figure~\ref{fig:teaser}B depicts this focal
\emph{target-free half} configuration with the AI-referenced location
still hidden.

\paragraph{Focal contextual comparisons}
Our baseline condition had 50\% target prevalence and a fully hidden
desk.
Relative to this baseline, the focal analyses examine three changes in
context:
(1) \textbf{stronger prior}, with 25\% prevalence and a fully hidden
desk;
(2) \textbf{target-free half}, with 50\% prevalence and no key in
the visible half; and
(3) \textbf{combined context}, with 25\% prevalence and no key in the
visible half.
In all three comparisons, the target state remained unresolved from the
participant's view.

\subsection{AI Assistance}

We used simulated AI advice so that participants received a controlled
mixture of correct and incorrect suggestions.
Each 40-trial AI-assisted block contained 32 correct and eight incorrect
suggestions, giving the AI 80\% accuracy.
Errors were distributed proportionally across present- and absent-target
trials, so the AI was 80\% accurate for both target states within each
full block.
An incorrect suggestion therefore predicted \emph{Absent} when a key
was actually present, or \emph{Present} when no key was present.
Motivated by evidence that stated model accuracy can shape trust~\cite{Yin2019AccuracyTrust}, participants knew that the AI could be wrong but were not told its numeric accuracy.

When a key was present, a correct \emph{Present} suggestion referred to
its row. False-positive suggestions occurred only on absent-target
trials; they referred to the hidden row at 0\% visibility and could
refer to either row at 50\% visibility. Thus, finding the indicated
region target-free contradicted the spatial claim. This observation
alone would not generally establish absence from the whole desk; in
our simulation, false-positive spatial claims coincided with desk-wide
absence.

\paragraph{Presentation formats}
The full experiment included one \noai{} condition and three ways of
presenting the same AI advice: \verbal{}, \arrow{}, and \xray{}.
The \noai{} condition measured decisions without AI assistance.
In \verbal{}, audio reported \emph{Present} or \emph{Absent} and, at
50\% visibility, provided a coarse region for positive suggestions.
In \arrow{}, a spatial arrow indicated a positive prediction; a
negative prediction displayed the text \emph{Key is absent}
~\cite{Lee2023DesignPatterns,Liu2026ARReliance}.
In \xray{}, a transparent view exposed the desk behind the divider.
A translucent key at the predicted location represented a
\emph{Present} suggestion
~\cite{Livingston2003Occlusion,Elmqvist2008OcclusionTaxonomy,Liu2026ARReliance}.
For an \emph{Absent} suggestion, the view showed no key, including
false-negative trials on which the task target was present. Thus,
although the desk behind the divider was shown, the x-ray key display
encoded the AI prediction rather than independently revealing the
hidden target state. Figure~\ref{fig:trial_process} shows the formats,
response controls, and feedback.

The pooled analyses estimate contextual effects averaged across these formats; format comparisons are outside the focal research questions.

\subsection{Participants and Apparatus}

We recruited 32 participants (16 men and 16 women): 27 were aged
18--28, four were 29--38, and one was 39--48.
Recruitment used university email, word of mouth, and posted fliers.
Participants met vision, color-vision, and motion-sickness eligibility
criteria, completed a 75-minute session, and received \$25--30.

Prior headset experience was limited: two participants reported no prior
use, 19 reported 0--10 uses, seven reported 11--50 uses, and four
reported more than 50 uses.

The study ran on a Meta Quest 3 using a Unity client aligned to scanned
desk geometry.
A server controlled the target row and experimental condition, while
the headset client selected the target's horizontal position and spatial
jitter within the assigned row.

\subsection{Procedure}

Each participant completed 16 counterbalanced blocks:
two prevalence levels $\times$ two visibility levels $\times$ four AI
conditions, with 40 trials per block.
This yielded 640 planned trials per participant and 20,480 planned main
trials overall.
One trial was missing from the recorded data, leaving 20,479 recorded
trials, of which 20,456 contained an answer.

Participants first completed a 10-minute tutorial followed by 32
training trials covering both visibility levels, all AI conditions,
both target states, and correct and incorrect AI advice.
They then completed the 16 main blocks.
The interface separately displayed the current prevalence, visibility, AI
condition, trial number, and block score.

We recorded each final judgment and decision time.
After each response, participants saw the correct answer and a score
change. Figure~\ref{fig:trial_process} summarizes this
view--respond--feedback sequence.
They completed a short questionnaire after each AI-assisted block and
could rest between blocks.
A five-minute semi-structured interview followed the main task.

\begin{table*}[t]
  \centering
  \small
  \caption{Overview of the main quantitative findings.}
  \label{tab:results-overview}
  \begin{tabular}{@{}p{0.17\textwidth}p{0.38\textwidth}p{0.39\textwidth}@{}}
    \toprule
    Question & Main finding & Interpretation \\
    \midrule

    \textbf{Q1: Accuracy}
      & The \emph{combined context} increased observed AI-assisted accuracy by
        9.5 percentage points.
      & Most of the gain was attributed to the scored target and AI-suggestion mix; changed responses did not significantly improve accuracy beyond the baseline-response estimate. \\

    \addlinespace

    \textbf{Q2: Responses to AI}
      & The \emph{stronger prior} shifted responses away from \emph{Present}
        suggestions. Agreement fell significantly for correct \emph{Present}
        suggestions; the decrease for incorrect \emph{Present} suggestions was nonsignificant.
      & We did not detect improved agreement selectivity. A response-policy shift does not by itself establish better or worse use of available evidence. \\

    \addlinespace

    \textbf{Q3: Decision time}
      & The \emph{target-free half} increased decision time,
        both with and without AI.
      & Longer decision time was not accompanied by a detected selectivity gain; inspection was not measured separately. \\

    \bottomrule
  \end{tabular}
\end{table*}

\subsection{Analysis Scope and Measures}

The focal analyses retain fully occluded trials and half-visible trials
with a \emph{target-free half} (Fig.~\ref{fig:context-arrangement}).
For AI-assisted trials, we additionally exclude suggestions referring
to the visible row, where the spatial claim could be directly checked
(Fig.~\ref{fig:teaser}A). Retained trials leave both target presence
and any positively indicated location unresolved
(Fig.~\ref{fig:teaser}B). Removing visible targets and visible-row
false positives changes both the target mix and the accuracy of
retained AI suggestions. The AI and \noai{} subsets therefore differ.

Our three primary measures map directly onto the research questions:

\begin{itemize}[noitemsep,leftmargin=*]
    \item \textbf{Decision accuracy} is whether the participant's final
    \emph{Present}/\emph{Absent} response matched the ground truth.

    \item \textbf{Responses to AI suggestions} measure how often
    participants followed correct and incorrect suggestions.
    \emph{Agreement selectivity} is following of correct minus incorrect
    suggestions, calculated separately for \emph{Present} and
    \emph{Absent} suggestions.

    \item \textbf{Decision time} is the time from trial onset to the
    participant's final response.
\end{itemize}

Because participants did not provide a pre-advice judgment, following
measures final agreement with the AI rather than advice-induced
switching
~\cite{Schemmer2023AppropriateReliance,Wang2021ExplanationsHelpful,
Fok2024Verifiability}.

Additional in-task probes collected confidence ratings with and without
AI and trust ratings with AI, using seven-point scales. These secondary
measures remain in the correction families below but are outside the
focal research questions.

\subsection{Statistical Analysis}
\label{sec:statistical-analysis}

We compared each contextual condition with baseline and used an
exploratory accuracy analysis to distinguish changes in participant
responses from changes in the cases being evaluated.

\paragraph{Baseline-response estimate and decomposition}
For each participant and AI presentation, we estimated baseline response
accuracy in the four target-state-by-suggestion cells. Applying these
rates to a comparison condition's cell proportions gives its
\emph{baseline-response estimate}: expected accuracy if baseline response
patterns were maintained on those trials. Observed accuracy minus this
estimate is the \emph{response-change component}.

The decomposition further divides the change accounted for by trial
composition. Holding baseline cell accuracies fixed, we separate target
proportions from suggestion probabilities conditional on target state.
Each contribution is averaged over both replacement orders for these
two factors. Together with response change, the components sum exactly
to observed minus baseline accuracy. Components are averaged equally
across presentations before participant-level inference. This is a
descriptive reference, not a causal estimate or an optimal policy.

\paragraph{Decision-time preparation}
Timing analyses used answered trials with recorded times $0<t\leq20$~s,
excluding 42 longer responses and 17 missing times from the focal set.
We analyzed log times, reported geometric means, and transformed
log contrasts to percentage changes as $100[\exp(\Delta)-1]$.

\paragraph{Participant-level inference}
We calculated condition means within participant and presentation,
then weighted the three AI presentations equally within participant.
Two-sided paired $t$-tests compared each contextual condition with
baseline, separately with and without AI. Differences from baseline-response
estimates and decomposition components used two-sided one-sample $t$-tests
against zero. Participant-level estimates have unadjusted 95\%
$t$-intervals; significance uses Holm-adjusted $p_{\mathrm{Holm}}<.05$.

\paragraph{Multiple comparisons}
Holm correction was applied separately to 27 main tests
(three comparisons for each of nine outcomes: accuracy, decision time,
and confidence with and without AI; following correct advice; rejecting
incorrect advice; and AI trust), and 42 follow-ups (18 baseline-response
comparisons, 18 target/suggestion-specific contrasts, and six selectivity
contrasts). These families retain all tests, including outcomes not
discussed here. The exploratory decomposition forms a separate nine-test
family (three contexts $\times$ three components).

\paragraph{AI versus no-AI timing comparison}
To test whether context changed decision time differently with and
without AI, we fitted a Gaussian generalized estimating equation model
of log time with context-by-AI interactions. It adjusted for target
presence and standardized within-block trial position and block order,
using participant clusters, exchangeable correlation, and robust standard
errors. Each participant/context/assistance cell received equal total
weight: individual trial weights were inversely proportional to that
cell's trial count. Two-sided Wald tests of the three interaction
coefficients were Holm-corrected as a separate family. These compare
proportional timing changes across the retained subsets.

\begin{figure*}[htb]
    \centering
    \includegraphics[width=\textwidth]{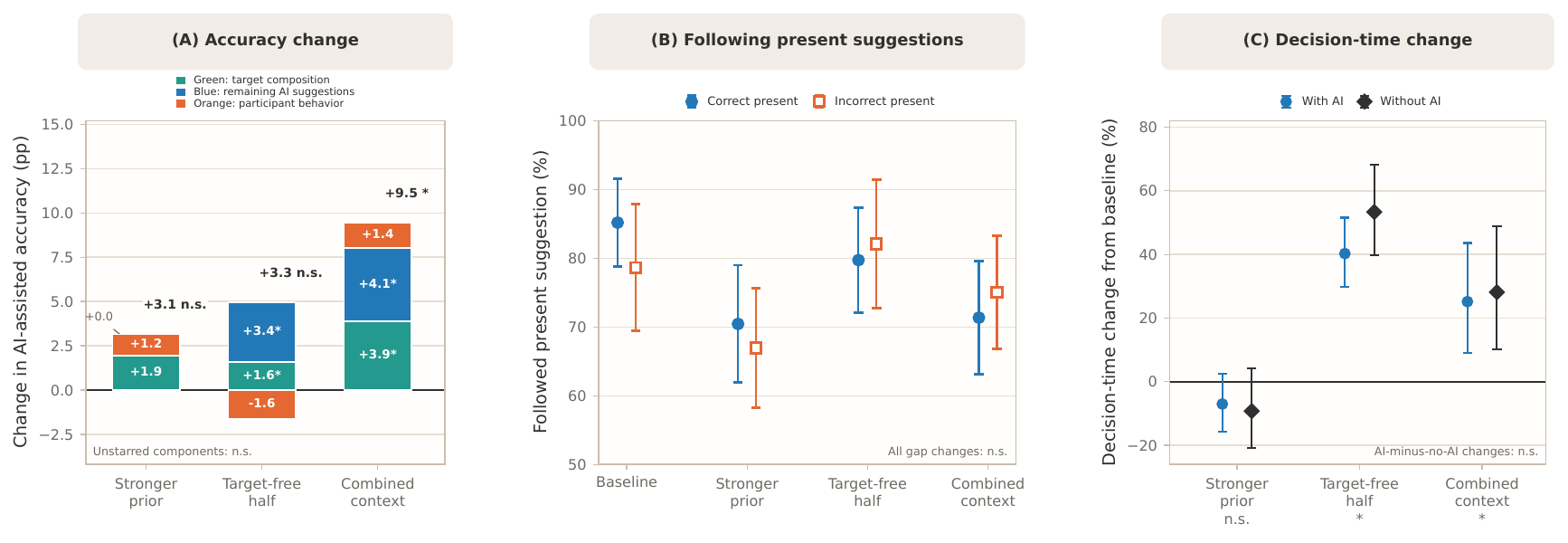}
    \caption{Context effects on focal unresolved trials. (A) AI-assisted accuracy change from baseline, split into target composition, retained-suggestion accuracy, and response change. Stars on totals use the 27-test family; component stars use the nine-test family. (B) Following of correct and incorrect \emph{Present} suggestions; ``gap changes'' are changes in correct-minus-incorrect following (42-test family). (C) Geometric decision-time changes; stars compare each assistance condition with baseline (27-test family). AI-minus-no-AI interactions use a separate three-test family and are nonsignificant. Error bars are unadjusted 95\% $t$-intervals; stars denote $p_{\mathrm{Holm}}<.05$.}
    \label{fig:aixvr-results-overview}
\end{figure*}

\section{Results}
\label{sec:results}

\subsection{Quantitative Results}
\label{sec:quantitative-results}

The analysis included all 32 participants, with 13,418 answered
AI-assisted trials and 4,637 answered \noai{} trials.
The three contextual comparisons use the fully hidden, 50\%-prevalence baseline defined in Section~\ref{sec:design}.
Table~\ref{tab:results-overview} summarizes the three main findings.
Unless otherwise noted, tests use participant-level condition means
with Holm-corrected $p$ values.

% ============================================================
\subsubsection{Q1: Does Context Improve Decision Accuracy?}
\label{sec:res-accuracy}

\paragraph{Observed accuracy increased significantly with AI only when the combined context was present}
At baseline, AI-assisted accuracy was 73.7\%.
The stronger prior alone increased accuracy by $+3.1$~pp
($p_{\mathrm{Holm}}=.16$), and the target-free half
increased it by $+3.3$~pp ($p_{\mathrm{Holm}}=.34$); neither change was
significant.
With the combined context, accuracy increased to 83.1\%
($+9.5$~pp, $p_{\mathrm{Holm}}<.001$).

Without AI, all three changes increased accuracy.
Accuracy rose from 49.9\% at baseline to 66.3\% under the stronger prior
($+16.4$~pp), 57.8\% with the target-free half ($+7.9$~pp), and
78.4\% with the combined context ($+28.5$~pp).
All three increases were significant ($p_{\mathrm{Holm}}\leq.010$).

\paragraph{Accuracy relative to baseline response behavior}
Using the exploratory baseline-response estimate defined in
Section~\ref{sec:statistical-analysis}, we compared observed accuracy
with accuracy expected from unchanged baseline responses on the new
trial distribution.
Observed accuracy differed from this estimate by $+1.2$~pp under
the stronger prior, $-1.6$~pp with the target-free half, and
$+1.4$~pp with the combined context; none of these differences
was significant.

\paragraph{Exploratory decomposition of the observed gain}
The decomposition extends the preceding baseline-response comparison
by splitting its trial-composition contribution into target mix and
retained AI suggestions (Fig.~\ref{fig:aixvr-results-overview}A).
Its response-change component is the same observed-minus-reference
difference reported above.
When the stronger prior and the target-free half were combined,
$+3.9$~pp of the accuracy gain was accounted for by the target mix
($p_{\mathrm{Holm}}=.008$), and another $+4.1$~pp by the higher
accuracy of the AI suggestions remaining in the analysis
($p_{\mathrm{Holm}}<.001$).
The response-change component was $+1.4$~pp
($p_{\mathrm{Holm}}=.879$).

The same pattern appeared when only the target-free half was present:
$+1.6$~pp was accounted for by the target mix
($p_{\mathrm{Holm}}=.009$), $+3.4$~pp from AI accuracy
($p_{\mathrm{Holm}}<.001$), and $-1.6$~pp from changed participant
responses ($p_{\mathrm{Holm}}=.879$).
Under the stronger prior alone, the corresponding components were
$+1.9$, $+0.0$, and $+1.2$~pp; none was significant
(all $p_{\mathrm{Holm}}\geq.110$).

% ============================================================
\subsubsection{Q2: How Does Context Change Responses to AI Suggestions?}
\label{sec:res-response}

\paragraph{The stronger prior shifted agreement with \emph{Present} suggestions}
Following of correct \emph{Present} suggestions fell from 85.2\% at
baseline to 70.4\% under the stronger prior
($-14.7$~pp, $p_{\mathrm{Holm}}<.001$).
Following of incorrect \emph{Present} suggestions decreased numerically
from 78.6\% to 66.9\%, but this contrast was nonsignificant
($-11.7$~pp, $p_{\mathrm{Holm}}=.075$).
Thus, the stronger prior (lower prevalence) made participants less willing to accept
a \emph{Present} answer in general, including when the AI was correct.

Across \emph{Present} and \emph{Absent} suggestions combined,
rejection of incorrect suggestions increased from 14.7\% to 25.3\%
($+10.6$~pp, $p_{\mathrm{Holm}}=.011$).
Across these same two suggestion directions, following of correct
suggestions changed by only $+1.3$~pp
($p_{\mathrm{Holm}}=1$).

\paragraph{No detected change in agreement selectivity}
To distinguish this general response shift from selective agreement,
we returned to each suggestion direction separately and compared
following of correct versus incorrect advice. Figure~\ref{fig:aixvr-results-overview}b reports the
\emph{Present}-suggestion rates for all four contexts; in the target-free
half and combined context, correct and incorrect rates remained close.
The six inferential contrasts included both suggestion directions; the
\emph{Absent}-suggestion rates are not plotted to keep the panel legible.
None of these correct-versus-incorrect differences changed significantly
in any of the three contexts (all six $p_{\mathrm{Holm}}\geq.66$).

% ============================================================
\subsubsection{Q3: Does Context Increase Decision Time?}
\label{sec:res-time}

\paragraph{The target-free half made decisions slower, both with and without AI}
With AI, decision time increased from 1.38~s at baseline to 1.94~s
with the target-free half
($+40.3\%$, $p_{\mathrm{Holm}}<.001$)
and to 1.73~s with the combined context
($+25.2\%$, $p_{\mathrm{Holm}}=.038$).
The stronger prior alone did not significantly change decision time
($-7.0\%$).

Without AI, decision time similarly increased from 1.08~s to 1.66~s
with the target-free half
($+53.3\%$, $p_{\mathrm{Holm}}<.001$)
and to 1.39~s with the combined context
($+28.1\%$, $p_{\mathrm{Holm}}=.037$).
The stronger prior alone again produced no significant change
($-9.2\%$).

\paragraph{No detected difference in proportional slowdown with versus without AI}
The trial-level context-by-AI interactions were nonsignificant
($p_{\mathrm{Holm}}\geq.286$; separate three-test family;
Fig.~\ref{fig:aixvr-results-overview}C).
This pattern is consistent with additional scene inspection, but elapsed
decision time does not identify the activity responsible for the slowdown.

%%%%%%%%%%%
\begin{samepage}
\subsection{Qualitative Results}
\label{sec:qualitative-results}

We used reflexive thematic analysis of open-text responses from 381
post-block questionnaires and semi-structured interviews with all 32
participants. Concepts of visibility, prevalence, and AI reliance
sensitized the reading; themes were developed by comparing shared
meanings across participants and conditions. We compared these themes
with the quantitative results and used divergent accounts to qualify
their scope, including reports of increasing, decreasing, and unchanged
reliance. These accounts describe participants' interpretations of their
strategies, rather than establishing what caused individual responses.
Because questionnaire submissions were uneven, thematic coverage was
considered at the participant level rather than by response frequency.

\end{samepage}

\subsubsection{Indirect Context Became a General Decision Policy}

Participants described using prevalence to decide which response was
plausible. In a 25\% block, P05 wrote, ``I mainly used the 25\% odds,'' while
P12 described trying to answer ``by every 4.'' When the relevant part of the
scene remained hidden, participants often treated the AI as the only
trial-specific cue. P04 explained, ``I always chose what AI said because I
had no information outside of this.'' These accounts suggest that
participants used indirect context as a general response tendency rather than
as evidence for resolving an individual AI claim.

\subsubsection{Direct Evidence Created an Override Boundary}

Participants spoke differently about evidence that could settle the current
claim. P14 listened to the AI first but said, ``if I could prove it wrong, I
would do what I could see.'' P04 similarly followed visible evidence when a
key or an indicated empty location could be checked, but otherwise followed
the AI. Participants also described scanning and checking the scene, consistent
with the longer decision times in half-visible contexts. These accounts
illustrate how participants could use directly checkable observations to
evaluate an AI's specific claim, whereas the contextual information in the
focal trials only changed its plausibility.

\subsubsection{Reliance Followed Subjective Models of the AI}
Participants described updating their reliance based on recent AI performance.
P26 followed the AI after several correct suggestions but trusted it less when
the streak broke. The direction was not uniform: some participants reported
increasing reliance, others decreasing it after visible mistakes, and others
no change. These accounts suggest that participants formed subjective models
of AI reliability from local experience rather than following a uniform
reliance strategy.

\section{Discussion}
\label{sec:discussion}

Context changed participants' decisions while the focal trials left the relevant evidence hidden. The response shift under the \emph{stronger prior}, the slowdown with a \emph{target-free half}, and the changing composition of evaluated cases point to different consequences of providing contextual information.

\subsection{Context Shapes a Decision Policy}
\label{sec:discussion-plausibility}

Participants' accounts connect the effect of the \emph{stronger prior} to a general decision policy. They described using the stated odds to guide their answers, overriding AI when visible evidence contradicted it, and adjusting reliance after recent successes or mistakes. These accounts suggest that participants combined expectations about the environment with subjective estimates of AI reliability. When the relevant scene remained hidden, the AI could still be their only trial-specific source of information.

The benchmark in Section~\ref{sec:rationale-context} gives a concrete reason to interpret these policies carefully: lower prevalence changes posterior probability but does not change the best binary answer when the programmed AI reliability is known. Participants had to learn that reliability through use, making their reported beliefs relevant when interpreting response shifts~\cite{Fok2024Verifiability}.

\subsection{Scene Visibility and Claim Verification}
\label{sec:discussion-evidence}

AR makes the distinction between locating a claim and checking it particularly concrete. A spatially registered arrow or virtual object can identify where the AI predicts a key, even when that location lies behind an occluder. The cue's position connects the prediction to the physical scene, while evidence about the predicted object remains inaccessible (Fig.~\ref{fig:teaser}).

The target-free half exposed more of the scene and increased decision time, but left the AI-referenced location hidden in the focal trials. The longer decision time was not accompanied by a detected increase in agreement selectivity. Participants' descriptions of scanning are consistent with inspection, but timing does not distinguish inspection from deliberation or other response processes. Their accounts of overriding AI with visible contradictions suggest why access to the specific claim-relevant location matters. Liu et al. show that perceptual difficulties and embedded representations can shape reliance in AR~\cite{Liu2026ARReliance}; our findings highlight the complementary issue of which physical evidence the interface allows users to inspect. This motivates distinguishing the amount of visible space from its relevance to the current prediction.

\subsection{Accuracy Changes with the Unresolved Case Set}
\label{sec:discussion-accuracy}

The exploratory decomposition shows how a higher score on unresolved cases can be accounted for largely by their composition (Fig.~\ref{fig:aixvr-results-overview}A). Revealing part of a scene changes which decisions remain unresolved, so the evaluated cases become an outcome of evidence access. This matters whenever an interface changes what users can verify: apparent performance gains on the remaining cases should be interpreted alongside changes in the target and AI-suggestion distributions.

\subsection{Design Implications}
\label{sec:discussion-design}

We examined how people combine target prevalence, partial visibility, and fallible AI advice when deciding whether a target is present. The findings suggest that situated AI interfaces should help users recognize what they can verify and weigh advice when direct evidence is unavailable. They also show why evaluations need to account for which decisions remain unresolved as the interface changes what users can see.

\paragraph{Make unresolved AI claims explicit}
An AR cue can point to a predicted object even when users cannot inspect that location from their current viewpoint. Interfaces should distinguish what users can observe directly, what the AI infers, and what remains uncertain. For example, an assistant could indicate that a predicted object is behind an occluder or that the claim is based on an earlier observation or another camera. Showing whether a claim is currently checkable, separately from the AI's confidence, may help users recognize when they are acting on a prediction.

\paragraph{Help users weigh AI advice against what they know about the environment}
In a partially visible environment, people can draw on how often a target usually appears, what they can see, and what the AI predicts on the current trial. In our study, lower target prevalence made participants less likely to follow even correct \emph{Present} advice. Participants also described adjusting their reliance after seeing the AI succeed or fail. An AR assistant could present the target's base rate, the AI's track record, and its current prediction as distinct cues. Future work could test whether separating these cues helps users weigh advice when direct evidence is unavailable.

\paragraph{Account for which decisions remain unresolved when the interface changes what users can see}
Revealing part of a scene can make some claims directly checkable, changing the mix of cases in which users still have to judge AI advice without direct evidence. In our study, an exploratory analysis attributed most of the observed accuracy gain on unresolved trials to changes in the mix of target-present and target-absent cases and in the accuracy of the AI advice on those trials. Evaluations should report how many cases become checkable and how people perform on the task overall. To assess whether the interface also changes how people use AI advice, researchers should compare unresolved trials with similar distributions of target states and AI suggestions.

\subsection{Limitations and Future Work}
\label{sec:discussion-limitations}

The study involved 32 mostly young adults, one binary search task with a Unity-rendered target registered to a physical desk, and simulated AI with 80\% full-block accuracy. Transfer to searches for physical objects remains to be tested. Prevalence was displayed, whereas AI reliability had to be learned. The analytical benchmark assumes the programmed reliability and equal error costs; it does not model participants' subjective beliefs or trial history. Longer-term use, other reliability levels, and different error costs may support different policies. Pooling presentation formats estimates their average contextual effect and may obscure format-specific differences.

The focal analyses excluded directly checkable trials, changing the evaluated distribution and limiting the opportunity to distinguish correct from incorrect claims using visible evidence. The baseline-response comparison and decomposition are exploratory descriptions, rather than tests of optimal policy use or causal estimates. The AI and \noai{} timing subsets also differ because only AI trials exclude visible-row suggestions; their comparison cannot isolate AI processing costs. Nonsignificant contrasts do not demonstrate equivalence or exclude useful effects. Prospectively balanced unresolved trial sets would provide a stronger basis for estimating behavioral changes.

Finally, post-advice binary judgments cannot distinguish independent agreement from advice-induced switching or isolate unaided perception from AI-supported judgment. Pre-advice judgments, probability estimates, and inspection traces could connect belief updating and evidence acquisition to final choices. Combining these measures with Bayesian and signal-detection approaches~\cite{Manavalan2026VisibilityPrevalence} would help evaluate attainable policies using the joint information in the scene and AI cues.

\begin{samepage}
\section{Conclusion}
\label{sec:conclusion}

We studied whether contextual information from a partially visible physical environment helps people make better decisions with fallible AI. The \emph{stronger prior} (lower prevalence) made participants more skeptical of \emph{Present} suggestions, and the \emph{target-free half} increased decision time. On unresolved trials, we detected no increase in selective agreement with correct versus incorrect advice. An exploratory analysis attributed most of the observed accuracy gain to changes in the cases evaluated, without a demonstrated gain from changed participant responses.

Imagine searching for a tool in a workshop. An AR assistant predicts that it is behind a cabinet. The interface could indicate that the location is hidden, suggest a viewpoint from which to inspect it, and show when the claim becomes checkable. When no direct view is available, it could help the user weigh the prediction by distinguishing what the AI predicts now, how it has performed in similar searches, and how often the tool is found there.

Future studies could compare guidance toward two viewpoints that reveal equally large regions, with only one exposing the predicted location. Evaluations should measure accuracy and decision time across the task, track which cases become checkable, and compare unresolved cases with similar mixes of targets and AI suggestions. They could also test what users carry into an unfamiliar scene. Once the guidance is removed, can people still identify what to inspect to settle a new claim?
\end{samepage}

\bibliographystyle{IEEEtran}
\bibliography{ref}

\end{document}